
\magnification=1200
\vsize=22truecm
\hsize=15truecm \tolerance 1000
\parindent=0pt
\baselineskip = 15pt
\lineskip = 1.5pt
\lineskiplimit = 3pt
\voffset=0pt

\def\ket#1{\vert #1 \rangle}
\parskip = 1.5ex plus .5ex minus .1ex
{\nopagenumbers
\rightline{KCL-TH-93-10}
\vglue 1truein
\centerline{$W_3$  STRINGS, PARAFERMIONS AND THE ISING MODEL}
\bigskip
\centerline{Michael Freeman}
\medskip
\centerline{Department of Mathematics}
\centerline{King's College London}
\centerline{Strand, London WC2R 2LS}
\centerline{\&}
\centerline{Peter West}
\medskip
\centerline{Department of Mathematics}
\centerline{King's College London}
\centerline{Strand, London WC2R 2LS}
\centerline{\&}
\centerline{Institute of Theoretical Physics}
\centerline{Chalmers University of Technology}
\centerline{S-412 96 G\"oteborg SWEDEN}
\bigskip
\centerline{June 1993}
\vskip .75truein
\centerline{Abstract}
\smallskip
We show that any covariant scattering amplitude of the $W_3$ string will
contain, as part of its integrand, a factor that obeys the differential
equations satisfied by an Ising model
correlation function.  This factor can thus be identified with such a
correlation function, in agreement with a previous result of the
authors.  The $W_3$ string is also shown to contain an $N=2$ parafermion
theory, and hence to contain
in addition the non-linear infinite-dimensional $W$-algebra
corresponding to this parafermion theory.  The physical states form a
representation of this algebra.
\vfil
\eject}

{\bf Introduction}

One of the most remarkable features to emerge from the development of
$W_N$ strings is their relation to the unitary minimal models.  The first
suggestion that such a relation existed arose from the observation of
a numerical coincidence [1,2], namely that the central charge of a
subset of the ghost and matter fields of a $W_N$ string was the same as
that of a unitary minimal model.  The simplest example of this is given
by the $W_3$ string, which is related to the Ising model.  This string can
be constructed from scalar fields $\phi$ and $X^\mu$, with $\mu = 0,1,
\ldots, D-1$, together with reparametrization ghosts $b,c$ and
ghosts $d,e$ corresponding to $W_3$-transformations.  It can then be seen
that the fields $\phi$, $d$ and $e$ have a total central charge of $1/2$,
which is precisely that of the Ising model.  Furthermore, the intercepts
associated with the $X^\mu$-sector of physical states of the $W_N$ string
are related to the weights of primary fields of the corresponding
minimal model [1,2].

More recently, direct connections between $W_N$ strings and the unitary
minimal models have been discovered.  The partition function of the
$W_3$ string, which gives the count of states at each level, has been
shown to involve the characters of the Ising model [3].  In addition, the
tree-level scattering amplitudes for the $W_3$ string contain Ising
model correlation functions, and satisfy the factorization and
duality properties expected of a string theory [4].

In order to explain these connections in more detail, let us review some
facts about $W_3$ strings.  The physical states of the $W_3$ string theory
are given by the cohomology of the BRST charge $Q$.  It has been found [5]
that, with the exception of certain discrete states, this cohomology can
be obtained by the action of a picture-changing operator $P$ and a
screening charge $S$ on three distinct states $\ket{\psi_a}$, with
$a=1$, $15/16$ and $1/2$.  For $a=1$ and $15/16$, the states
$\ket{\psi_a}$ have the form
$$
\eqalign{
\ket{\psi_a} &= c\, \partial e \, e\, e^{i \beta(a,0) \phi} V^X(a) \ket{0}\cr
&\equiv V(a,0) \ket{0},\cr}
\eqno(1)
$$
while $\ket{\psi_{1/2}}$ has the form
$$
\eqalign{
\ket{\psi_{1/2}} &= (c\, e - {i\over\sqrt{522}}\partial e\,e)
e^{i\beta(1/2,0)\phi} V^X(1/2) \ket{0}\cr
&\equiv V(1/2,0) \ket{0};\cr}
\eqno(2)
$$
here $\ket{0}$ is the invariant vacuum state, $V^X(a)$ is a conformal operator
of weight $a$ constructed from $X^\mu$ alone, and the momenta $\beta$
are given by
$$
\beta(1,0) = 8iQ/7, \quad \beta(15/16,0) = iQ,\quad \beta(1/2,0) =4iQ/7.
\eqno(3)
$$
The screening operator $S$, which is used to generate further states in the
cohomology, is given by [5]
$$
S=\oint dz\left (d-{5i\over 3\sqrt{58}}\partial b -
{2\over 261}\partial b b e - {41\over 3\sqrt{58}}d b e\right) e^{i\beta_s\phi}
\eqno(4)
$$
with $\beta_s = -2iQ/7$.

We have seen that, with the exception of the discrete states, the physical
states are encoded in $V^X(a)$.  The corresponding count of states has been
shown [3] to be given by the character
$$
\left\{\prod_{n=1}^\infty {1\over (1-x^n)^{D-2}}\right\} \chi_h(x),
\eqno(5)
$$
where $\chi_h$ is the Ising model character for the states of weight
$h=1-a$.

Following the general tree-level
calculations of $W_3$ string scattering amplitudes
given in [4], covariant approaches to $W_3$ string scattering have been
considered.  The first such attempt [6,7] involved calculations of
correlation functions of only vertex operators and picture-changing
operators, and led to results that disagreed with those of reference [4].
These amplitudes did not share the connection with Ising model correlation
functions found in [4], and in addition they violated general principles of
S-matrix theory and string theory.  The use of vertex and picture-changing
operators was a straightforward generalization of previous covariant
string scattering calculations; in view of the expected role of $W$-moduli
it is perhaps not surprising that this simple generalization
is not adequate for $W$-strings.

In reference [5] a general covariant procedure for calculating $W_3$-scattering
was given.  This involved using not only the vertices $V(a,0)$ and the
picture changing operator $P$ but also the screening charge $S$ in the
computation of scattering amplitudes, and led to
amplitudes in agreement with those found earlier [4] and which satisfied the
general principles expected.
A number of specific amplitudes were evaluated explicitly
using this formalism in
reference [5].  It is the purpose of this paper to show quite generally that
all amplitudes calculated in this way contain correlation functions of
the Ising model.

In reference [7] a field redefinition that leads to a considerable
simplification of the physical states was given for the $W_3$ string.
The redefined fields are given by
$$
\eqalign{
\tilde c &= c + {7\sqrt{58}i\over 174}\partial e -
{8\over 261}b\,\partial e\, e - {4\sqrt{29} i \over 87} \partial\phi\, e\cr
\tilde b &= b\cr
\tilde e &= e\cr
\tilde d &= d + {7\sqrt{58}i\over 174}\partial b -
{8\over 261}\partial b\, b\, e + {4\sqrt{29} i\over 87} \partial \phi\, e\cr
\tilde \phi &= \phi - {4\sqrt{29} i \over 87} b\, e,\cr}
\eqno(6)
$$
whereupon the BRST charge $Q$ [8] takes the form $Q = Q_0 + Q_1$, with
$$
Q_0 = \oint j_0, \quad Q_1 = \oint j_1,
\eqno(7)
$$
where
$$
j_0 = \tilde c \left(T_X + T_{\tilde\phi} + T_{\tilde d \tilde e} +
{1\over 2} T_{\tilde b \tilde c}\right)
\eqno(8)
$$
and
$$
j_1 = -{4\sqrt{29}i\over 261} \tilde e \left(
2(\partial\tilde\phi)^3 + 6 Q \partial^2\tilde\phi\,\partial\tilde\phi
+{19\over 4} \partial^3\tilde\phi +
9 \partial\tilde\phi\, \tilde d\, \partial \tilde e +
3 Q \partial\tilde d\, \partial\tilde e \right).
\eqno(9)
$$
In $j_0$ the stress tensors for the redefined fields are of precisely the
same form as those for the original fields.
$Q_0$ and $Q_1$ then satisfy the relations
$$
Q_0^2 = Q_1^2 = 0, \quad \{Q_0,Q_1\} = 0.
\eqno(10)
$$
In the remainder of this paper we shall work with the redefined
fields, but for simplicity of notation we drop the tilde.

It is useful to note that there are two different types of ghost
number, $N_{b,c}$ and $N_{d,e}$, corresponding to the ghosts $b,c$ and $d,e$,
so that operators can be characterized by the pair $(N_{b,c}, N_{d,e})$.
Thus the fields $b$, $c$, $d$ and $e$ have ghost numbers $(-1,0)$,
$(1,0)$, $(0,-1)$ and $(0,1)$ respectively.  Then $Q_0$ and $Q_1$ have
ghost numbers $(1,0)$ and $(0,1)$, so that if $X$ is some operator that
commutes with $Q$ and which is an eigenstate of both $N_{b,c}$ and
$N_{d,e}$, then $X$ will commute with both $Q_0$ and $Q_1$.

In terms of the redefined fields, the basic vertices $V(a,0)$ are
$$
V(a,0) = c\, \partial e\, e\, e^{i\beta(a,0)\phi} V^X(a)
\eqno(11)
$$
for $a=0,15/16$, and
$$
V(1/2,0) = c\, e\, e^{i \beta(1/2,0)\phi} V^X(1/2).
\eqno(12)
$$
Further, the screening charge takes the particularly
simple form
$$
S=\oint dz \,  d\, e^{i\beta_s\phi}.
\eqno(13)
$$
Hence both $Q_0$ and $Q_1$ commute separately with the vertices
$V(a,0)$ and with $S$.

In general, however, the picture changing operator $P$ does not
commute separately with $Q_0$ and $Q_1$.  The most general form of
the picture changing operator is some linear combination of
$[Q,\phi]$ and $[Q, X^\mu]$.  An explicit calculation shows that
while $Q_0$ and $Q_1$ each commute with $[Q,X^\mu]$, the same is not
true of $[Q,\phi]$.  For our purposes, however, the important point to
note is that the only contribution of $P$ to a scattering amplitude comes
from the term $[Q_1,\phi]$.  This can be seen by considering the ghost
numbers of operators contributing to a scattering amplitude. An $N$ string
amplitude will contain 3 unintegrated vertex operators
$V(a,0)(z)$ together with $N-3$ integrated vertex operators
$\int dz \oint_z dw\,b(w) V(a,0)(w)$.  The total $b,c$ ghost number of
these operators is 3, which is precisely what is needed to get a non-zero
correlation function.  The scattering amplitude is obtained by including
insertions of $S$ and $P$, in addition to the physical state vertex operators.
Since $S$ has ghost number $(0,-1)$, and $P$ consists of terms having
ghost number either $(1,0)$ or $(0,1)$,
it follows that the only terms in $P$ that
can contribute are those of ghost number $(0,1)$.  These terms are
given precisely by $[Q_1,\phi]$, as stated above.  It is then evident that
$Q_1$ commutes with this part of $P$, and so $Q_1$ commutes with all of the
operators used in constructing a scattering amplitude for the $W_3$ string.
Furthermore, it can easily be seen that all correlation functions that arise
in this way will factorize into two separate correlation functions, one
involving the spacetime fields $X^\mu$ and the ghosts $b,c$, and the other
involving $\phi$ and the ghosts $d,e$.  Our aim is now to show that the
correlation function of $\phi$ and $d,e$ is, in fact, equal to a
correlation function of the Ising model.

At this point it is convenient to define some fields $\phi_h$ as follows:
$$
\phi_h = \cases{\partial e\, e\, e^{i\beta(1-h,0)\phi},&for $h=0,1/16$\cr
e\, e^{i\beta(1/2,0)\phi}&for $h=1/2$.\cr}
\eqno(14)
$$
The vertices $V(a,0)$ are then given in terms of $\phi_{1-a}$ by
$$
V(a,0) = c V^X(a) \phi_{1-a}.
\eqno(15)
$$
The fields $\phi_h$ are constructed only from the scalar field $\phi$ and
the ghost $e$.  It is straightforward to show that, with respect to the
stress tensor $T_\phi + T_{d,e}$ for these fields, $\phi_h$ is a primary
field with conformal weight $h$.  We have therefore constructed primary
fields of weights 0, $1/16$ and $1/2$ with respect to a stress tensor
having $c=1/2$, and we would like to show that the correlation functions
of these fields coincide with those of the Ising model.  The fact that the
weights of these fields are the same as those of the Ising model is not in
itself sufficient to show this; what is needed is to consider the
representations of the Virasoro algebra constructed from these fields.

It is well-known that the minimal conformal field theory models are solvable
as a consequence of the fact that there exist descendants of highest weight
states that are themselves highest weight states.  Such descendant states
can consistently be set to zero, and doing this leads to differential
equations that must be satisfied by correlation functions involving these
primary fields.  For the minimal models, these differential equations
enable all correlation functions of primary fields to be calculated.

Let us consider in more detail the $(p,q)$ minimal model, which has a
central charge $c=1-6(p-q)^2/pq$ and a closed operator product algebra
for a set of primary fields $\phi_{r,s}$ of weight
$h_{r,s} = [(rp-sq)^2-(p-q)^2]/4pq$, with $1\le r\le q-1$ and $1\le s \le p-1$.
The primary field $\phi_{r,s}$ has a descendant that is a highest weight field
at level $rs$.  Since it is possible to write $\phi_{r,s} = \phi_{q-r,p-s}$,
this field also has a highest weight descendant at level $(q-r)(p-s)$.
This field therefore possesses two highest weight descendants, and these
can be shown to be distinct in the sense that neither one can be expressed
as a descendant of the other.  Furthermore, all other highest weight
descendants in the Virasoro representation constructed from $\phi_{r,s}$ can
be written as descendants of these two fields.

The Ising model is obtained by taking $p=4$, $q=3$.  The three primary fields
of this model are $\phi_{1,1} = \phi_{2,3}$, with weight 0,
$\phi_{2,1}=\phi_{1,3}$, with weight $1/2$, and
$\phi_{1,2}=\phi_{2,2}$ with weight $1/16$.  Setting the highest weight
descendants of these fields to zero then gives the conditions
$$
\eqalign{
&\hat L_{-1}\phi_{1,1} =0\cr
&(\hat L_{-6} +{22\over 9} \hat L_{-4}\hat L_{-2} -{31\over 36}\hat L_{-3}^2
-{16\over 27}\hat L_{-2}^3)\phi_{1,1} =0\cr
&(\hat L_{-2} - {3\over 4}\hat L_{-1}^2) \phi_{2,1} =0\cr
&(\hat L_{-3} -{4\over 5} \hat L_{-1}\hat L_{-2} +
{4\over 15}\hat L_{-1}^3) \phi_{2,1} =0\cr
&(\hat L_{-2} - {4\over 3}\hat L_{-1}^2) \phi_{1,2} =0\cr
&(\hat L_{-4} -{122\over 147}\hat L_{-1}\hat L_{-3} +
{50\over 147}\hat L_{-1}^2\hat L_{-2}
-{4\over 49} \hat L_{-1}^4  - {1\over 36}\hat L_{-2}^2) \phi_{1,2} =0\cr
}
\eqno(16)
$$
Here we have used the notation $\hat L_n \phi(w) = \oint_w {dz\over 2 \pi i}
(z-w)^{n+1} T(z) \phi(w)$.

Given a particular realization of the fields of a $c=1/2$ conformal field
theory, it may or may not be the case that the above highest weight descendants
actually vanish.  Our method for showing the occurence of Ising model
correlation functions in $W_3$ string scattering amplitudes will be to
consider such highest weight descendants that can be constructed from the
vertex operators of the $W_3$ strings.  These descendants do not vanish
identically, but we will show that nevertheless they do give zero in
correlation functions.  This is sufficient to enable us to deduce the presence
of Ising model correlation functions in $W_3$ string scattering.

Let us therefore consider the highest weight descendants of the fields
$\phi_h$ given earlier.  The simplest possible example is
$\hat L_{-1}\phi_0$, which is equal to $\partial \phi_0/ \partial z$.
Evidently this is not zero, but it is simple to show that it can be
written as the commutator of $Q_1$ with some other operator.  The
obvious candidate for this operator is
$e e^{i\beta(1,0)\phi}$ and indeed we find
$$
{\partial \phi_0\over \partial z} =
{3\sqrt{29}\over 5 \sqrt 2}i\left\{Q_1, e e^{i\beta(1,0)\phi}\right\}.
\eqno(17)
$$
We have checked, using Mathematica [9] and the OPEdefs package of
Thielemans [10], that a similar picture holds for each of the other highest
weight descendants of the fields $\phi_h$ given in eqn (16); each such
highest weight descendant can be written as the commutator of $Q_1$ with
some primary field.  Thus $\phi_{1/2}$ satisfies the conditions
$$
\eqalign{
&(\hat L_{-2} - {3\over 4}\hat L_{-1}^2) \phi_{1/2} =
- {3\sqrt{29}\over 4\sqrt{2}} [Q_1,  e^{-4Q/7 \phi}]\cr
&(\hat L_{-3} -{4\over 5} \hat L_{-1}\hat L_{-2} +
{4\over 15}\hat L_{-1}^3) \phi_{1/2} =
- {3\sqrt{29}\over 10} [ Q_1, \partial\phi e^{-4Q/7 \phi} + {1\over\sqrt{2}}
d e e^{-4Q/7 \phi}],\cr
}
\eqno(18)
$$
and $\phi_{1/16}$ satisfies the conditions
$$
(\hat L_{-2} - {4\over 3}\hat L_{-1}^2) \phi_{1/16} =
-{\sqrt{29}\over 5} \{Q_1, {7\over\sqrt{2}} \partial e\,e^{-Q\phi} -
8 e \partial\phi\, e^{-Q\phi}\}
\eqno(19)
$$
and
$$
\eqalign{
&(\hat L_{-4} -{122\over 147}\hat L_{-1}\hat L_{-3} +
{50\over 147}\hat L_{-1}^2\hat L_{-2}
-{4\over 49} \hat L_{-1}^4  - {1\over 36}\hat L_{-2}^2) \phi_{1,2}\cr
&={9\sqrt{29}\over 4}
\Bigl\{Q_1,{745 Q\over 14112} d\,e\,\partial^2 e\,e^{-Q\phi} +
{25 Q\over 378} \partial d\,e\,\partial e\,e^{-Q\phi} -
{431 Q\over 15876} \partial^3 e \,e^{-Q\phi}\cr
&+{565\over 7056} d\,e\,\partial e \,\partial\phi\,e^{-Q\phi}+
{18071\over 42336}\partial^2e\,\partial\phi\,e^{-Q\phi} -
{103 Q\over 882} \partial e\,(\partial\phi)^2 e^{-Q\phi}\cr
&+{1\over 441}e (\partial\phi)^3 e^{-Q\phi}
-{29 Q\over 126} e \partial\phi\partial^2\phi\,e^{-Q\phi}
+ {23\over 63} \partial e\,\partial\phi\,e^{-Q\phi} +
{17\over 2016} e \partial^3\phi \, e^{-Q\phi}.\Bigr\}\cr}
\eqno(20)
$$
There is an analagous condition for the level-6 descendant of the field
$\phi_0$, but since this is lengthy we omit it here.

Since we have argued previously that $Q_1$ commutes with all operators
used to construct $W_3$ string scattering amplitudes, it follows that the
(anti-)commutator of $Q_1$ with any well-defined
operator will vanish in all correlation
functions of interest to us.  Hence all of the above highest weight
descendants can be set to zero, which implies that the correlation functions
of $\phi_0$, $\phi_{1/16}$ and $\phi_{1/2}$ satisfy certain
differential equations.  These differential equations, which are
sufficient to determine completely the correlation functions,
are precisely those
satisfied by correlation functions of the Ising model.  We therefore
conclude that
the integrand of any $W_3$ string scattering amplitude will contain a
factor that is an Ising model correlation function.

It was first observed [1,2] that, for certain low-level
physical states of the two-scalar $W_3$ string, one can regard the remaining
scalar outside the Ising sector as providing a Liouville dressing to the
Ising vertices.  It is clear from reference [5]  that this interpretation
holds for all vertices in the theory, so that the Ising vertices
$\phi_h$ are dressed in this way.  Since, as we have now shown, the Ising
sector vertices do indeed lead to Ising correlation functions as part
of the scattering amplitudes, it would be interesting to carry out
a detailed comparison between the full scattering amplitudes for the
$W_3$ string and those found for the Ising model coupled to 2-dimensional
gravity.

The $W_3$ string in fact contains an infinite-dimensional algebra that
elucidates the structure of the null states found above.  In order to
see this we consider the BRST charge $Q_1$ associated with the
Ising sector of the $W_3$ string. This charge can be used to
define a new spin-3 generator
$$
\eqalign{
U^{(3)}(z) \equiv& {9\sqrt{29} i\over 4} \{Q_1,d(z)\}\cr
=&2 (\partial\phi)^3 + 6 Q \partial^2\phi\,\partial\phi +
{19\over 4} \partial^3\phi + 18 \partial\phi\, d\,\partial e \cr
&+ 6 Q \partial d\, \partial e + 9 \partial\phi\,\partial d\, e +
9\partial^2\phi\, d\, e + 3 Q \partial^2d\, e.\cr}
\eqno(21)
$$
Both $U^{(3)}$ and the stress-tensor $U^{(2)}
\equiv T_\phi + T_{de}$ commute with
the BRST charge $Q_1$, but these fields do not form a closed algebra
amongst themselves.  The operator product of $U^{(3)}$ with itself can
be easily found from
$$
d(z) U^{(3)}(w) = - 6 {R^{(3)}(w)\over (z-w)^2} - 3 {\partial R^{(3)}(w)
\over (z-w)} + \ldots,
\eqno(22)
$$
where $R^{(3)} = 3 \phi d + Q \partial d$.  Taking the commutator of this
operator product with $Q_1$ then defines a new spin-4 generator
$U^{(4)}$ occurring in the operator product of $U^{(3)}$ with itself:
$$
U^{(3)}(z)U^{(3)}(w) = - 6 {U^{(4)}(w)\over (z-w)^2} -
3{\partial U^{(4)}(w)\over (z-w)} + \ldots.
\eqno(23)
$$
The operator product of $U^{(3)}$ with $U^{(4)}$ can be obtained in a
similar way by considering $R^{(3)}(z) U^{(4)}(w)$ and then taking the
anticommutator with $Q_1$.  The resulting OPE is found to be non-linear, and
contains a new spin-5 field $U^{(5)}$ in addition to $U^{(3)}$ and
$U^{(2)}$.  The existence of such an algebraic structure was first noticed
in reference [11] at the classical level, when considering under what
conditions a $W_N$ algebra would have a BRST charge with the decomposition
$Q= Q_0 + Q_1$ with $Q_0$ and $Q_1$ both squaring to zero.

It thus emerges that there is a non-linear infinite-dimensional algebra
associated with the Ising sector.  This algebra is generated by fields
$U^{(s)}$ of spin $s$, with $s \ge 2$, and $U^{(s)}$ is
of the form $U^{(s)} = \{ Q_1, R^{(s)}\}$ for $s \ge 3$.
It follows that the only
central term that arises in this algebra is in the OPE of $U^{(2)}$
with itself.

This infinite-dimensional algebra can be used to rewrite equations (17-20) in
terms of null vectors.  Let us consider $\phi_0$, for example.  This is
a highest weight state of the algebra, in the sense that
$$
\hat U^{(s)}_n \phi_0 = 0, \quad n \ge 1;
\eqno(24)
$$
where  $\hat U^{(s)}_n$ is defined by the operator product
expansion
$U^{(s)}(z) \phi(w) = \sum(z-w)^{-n-s}(\hat U^{(s)}_n\phi)(w)$.
Using the fact that $e e^{i \beta(1,0)\phi} = \hat d_{-1} \phi_0$,
we can rewrite
eqn (17) as the vanishing of a descendant field within this
infinite-dimensional algebra,
$$
(\hat U^{(3)}_{-1} + {15 Q \over 7}\hat L_{-1}) \phi_0 = 0.
\eqno(25)
$$
Similarly we see that $\phi_{1/16}$ is a highest weight field of the
infinite-dimensional non-linear algebra and that eqn (19) can be written as
$$
\left\{ 4 Q \hat L_{-2} -{4\over 3} \hat L_{-1}^2 + 17 \hat W_{-2}
-32 \hat L_{-1} \hat W_{-1}\right\}\phi_{1/16} = 0.
\eqno(26)
$$
A similar reformulation can be given for all of equations (17-20).

We now wish to investigate in more detail the structure of the
non-linear infinite-dimensional algebra discussed above.  To this end,
we shall show that it is a parafermion-generated $W$-algebra and identify
the parafermion currents in the Ising sector.  A parafermion theory [12]
contains currents $\psi_k, \psi_{-k} \equiv \psi^\dagger_k$,
for $ k = 0,\ldots, N-1$, with
both $\psi_k$ and $\psi^\dagger_k$ having conformal dimensions
$h_k=k(N-k)/N, \bar h_k = 0$.  $\psi_0$ is taken to be the identity operator.
The central charge of this theory is
$c=2(N-1)/(N+2)$, and so to make contact with the Ising sector with $c=1/2$
we choose $N=2$. $\psi_1$ and $\psi_{-1}$ then have conformal weight
$h=1/2$.  There are two operators in the cohomology of $Q_1$ that suggest
themselves for the roles of the parafermionic currents, namely
$$
\psi_1 = \phi_{1/2} = e e^{i \beta(1/2,0)\phi}
\eqno(27)
$$
and
$$
\psi_{-1} = \phi(1/2,1)
= \{6 \partial d\,d\,e + \sqrt{2} \partial\phi\partial d
-3\sqrt{2} \partial^2\phi d - 3/2\, \partial^2 d\} e^{i \beta(1/2,1)\phi},
\eqno(28)
$$
where $\beta(1/2,n) = (4 - 8n) i Q/7$ and $\phi(h,n)$ is defined by
$V(1-h,n) \equiv c V^X(1-h) \phi(h,n)$, $\phi(h,0) = \phi_h$.
If we bosonize the ghosts by
writing $d=e^{-i\rho}$ and $e = e^{i\rho}$, in which case their
energy-momentum tensor becomes
$T_{de} = -1/2(\partial\rho)^2 +5i/2\partial^2\rho$, we obtain
$$
\psi_1 = e^{i \rho + i \beta(1/2,0)\phi}
\eqno(29)
$$
and
$$
\psi_{-1} =
\{ -i \sqrt{2} \partial\phi\partial\rho - 3\sqrt{2}\partial^2\phi
+9i/2 \partial^2\rho - 3/2 (\partial\rho)^2\}e^{-i\rho + i\beta(1/2,1)\phi}.
\eqno(30)
$$
A straightforward calculation gives
$$
\psi_1(z)\psi_{-1}(w) = {1\over z-w}\sum_{n=0}^\infty (z-w)^n S_n,
\eqno(31)
$$
where
$$
S_n ={(n-1)(n-2)\over 2} P_n - (n-2) \partial P_{n-1} +
{1\over 2} \partial^2 P_{n-2} + 2 U^{(2)}P_{n-2}
\eqno(31)
$$
and
$$
P_n = e^{-i\rho - i\beta(1/2,0)\phi}
{\partial^n\over n!}e^{i\rho+i\beta(1/2,0)\phi}.
\eqno(33)
$$
Evaluating explicitly the right hand side, we find
$$
\eqalign{
& \psi_1(z) \psi_{-1}(w) = {1\over z-w}\times\cr
&\left\{1 + (z-w)^2 2 U^{(2)} + (z-w)^3 \partial U^{(2)}
+ {\sqrt{2}\over 3} (z-w)^3 U^{(3)} + O(z-w)^4 \right\},\cr}
\eqno(34)
$$
where $U^{(2)}$ is the total stress tensor for $\phi$ and $\rho$ and
$$
\eqalign{
U^{(3)} =& 2(\partial\phi)^3 + 6 Q \partial^2\phi\partial\phi
  + 19/4 \partial^3\phi \cr
&+ 9/2 \partial\phi\left((\partial\rho)^2 - 3i\partial^2\rho\right)
  - 9i\partial^2\phi\partial\rho \cr
&- Q \left(3\partial\rho\partial^2\rho + 2i\partial^3\rho + i(\partial\rho)^3
\right)\cr}
\eqno(35)
$$
is the bosonized form for $U^{(3)}$.

The first two terms in equation (34) confirm that $\psi_1$ and $\psi_{-1}$
are indeed parafermions.  It has been explained in references [13] how a
parafermionic operator product expansion contains in its regular part
an infinite number of terms which define the generators of an
infinite-dimensional non-linear $W$-algebra denoted $W_\infty(N)$ [13].
Eqns (31) and (35) show that
the algebra $W\infty(-2)$ generated by the parafermionic currents $\psi_1$ and
$\psi_{-1}$ is precisely the algebra discussed above.

It is known that the algebra $W_\infty(-2)$ constructed from the
parafermions $\psi_1$ and $\psi_1^\dagger$ has a realisation in terms
of two free fields [14].  It appears, however, that the realisation that
we have constructed here is inequivalent to that of reference [13]  This
can be seen by redefining the fields $\phi$ and $\rho$ so that only one
of them has a background charge;  although the stress tensor obtained in
this way agrees with that of reference [13], the spin-3 primary field does not.

In addition to containing parafermionic currents, the Ising sector also
carries a representation of the parafermionic algebra.  In the notation of
reference [12], the $N=2$ parafermion algebra can be represented on the fields
$\varphi^l_{[m]}$, where $l=0, 1$ and $m=-l, -l+2, \ldots, 4-l-2$.  The fields
$\varphi^l\equiv\varphi^l_{[l]}$ are parafermionic primary fields, in the sense
that they satisfy
$$
A_{l/2 + p}\varphi^l = A^\dagger_{-l/2 + p + 1}\varphi^l = 0, \quad p\ge 0,
\eqno(36)
$$
where $A_\nu$ and $A^\dagger_\nu$ are the modes of $\psi_1$ and
$\psi_{-1}$ respectively acting on $\varphi^l$.  The fields $\varphi^0$ and
$\varphi^1$ have dimensions $0$ and $1/16$ respectively.
The field $\varphi^0_0$ is the identity operator and can be identified with
$\phi(1,1)$.  The other primary field is $\varphi^1_1$, which we identify with
$e^{i\rho + i\beta(15/16,1)\phi}$.  It is straightforward to verify that this
is a parafermionic highest weight state.

The parafermions, and all of the above fields, commute with $Q_1$ since
they are defined in terms of $\phi(h,n)$ which itself commutes with $Q_1$.
It is also straightforward to show that $\psi_1$ and $\psi_{-1}$
commute with
the screening charge $S$, and hence all the generators of $W_\infty(-2)$
commute with both $S$ and $Q_1$.

The $W_\infty(-2)$ algebra is associated, through the underlying parafermions,
with a level 2 $SU(2)$ WZWN model, and one can use one of the other
scalar fields
of the $W_3$ string to realize the full $SU(2)$ currents rather than just the
$SU(2)/U(1)$ of the parafermions.  One can also incorporate $N=2$
supersymmetry into the model by a similar procedure.  Further results, and
details of the role of parafermions, will be given elsewhere [15].

The properties of the $W_N$ string theories for $N\ge 4$ are largely
unknown, although the numerical phenomenology of references [1,2] extends
to these theories.  It has also been found [3] that if these theories obey a
no-ghost theorem then their partition functions must involve the characters
of the unitary minimal models with $c=1-6/N(N+1)$.
There is evidence, however, that the $W_N$ string involves an $N-1$ parafermion
theory, thus generalizing the above results for the $W_3$ string.
A $W_N$ string theory involves scalar fields $\phi_j$, for
$j = 2, \ldots, N$, and ghosts $c_j, b_j$ for the same values of $j$.  The
field
$\phi_2$ can be replaced by any number of spacetime fields $X^\mu$ provided
they have the same central charge as $\phi_2$.  The final scalar $\phi_N$ has a
background charge such that its central charge is
$1 + 3(N-1)(2N+1)^2/(N+1)$, while the highest spin ghosts $c_N, b_N$
have a central charge $- 2 (6N^2 - 6N + 1)$.  Hence the
$\phi_N, b_N, c_N$ system has a total central charge of $2(N-2)/(N+1)$, which
is just that of an $N-1$ parafermionic theory.  It is thus natural to suppose
that a $W_N$ string theory can be written as a $W_{N-1}$ theory coupled to
an $N-1$ parafermion theory.  We recall that such a parafermion theory
generates the non-linear infinite dimensional algebra $W_\infty(-(N-1))$, which
contains the usual $W_{N-1}$ algebra together with higher spin generators with
no central term, which lends support to the above conjecture.
Clearly, the states of the $W_N$ string will provide a realisation of these
symmetries.  These conjectures extend and generalize those of reference [11].

It is also natural to suppose that the BRST charge for the $W_N$ string,
denoted $Q(W_N)$, can be written as
$$
Q(W_N) = Q_0(W_{N-1}) + Q_1^{(N-1)},
$$
where $Q_1^{(N-1)}$ is a function of the fields $\phi^N$, $c^N$ and $b^N$ and
is such that the spin-$N$ field in the $W_\infty(-N-1)$ algebra is given by the
anticommutator of $Q_1$ with $b^N$.  Assigning a separate ghost number to
$c^N, b^N$ we would conclude that $Q_0^2 = Q_1^2 = \{Q_0,Q_1\} = 0$.
Given the two boson realization of the $N-1$ parafermions, we can use it to
construct the BRST charge $Q(W_N)$ from $Q(W_{N-1})$ [15].

Acknowledgement

While this paper was being prepared for publication we received a paper [16]
by C. Hull which commented on the connection with the Ising model and which,
following reference [11], mentioned the presence of
a $W_\infty$ algebra in the $W_3$ string.  The nature of this algebra was
not clarified.
\vfil
\eject

{{\bf References}}
\parskip 0pt
\item{[1]} S. Das, A. Dhar and S. Kalyana Rama,
Mod. Phys. Lett. {\bf A6} (1991) 3055; Int. J. Mod. Phys.
{\bf A7} (1992) 2295.
\item{[2]} S. Kalyana Rama, Mod. Phys. Lett. {\bf A}6 (1991)3531.
\item{[3]} P. West, ``On the spectrum, no-ghost theorem and modular
invariance of $W_3$ strings,'' KCL-th-92-7, to be published in
Int. J. Mod. Phys.
\item{[4]} M. Freeman and P. West, Phys. Lett. {\bf B299} (1993) 30.
\item{[5]} M. Freeman and P. West; ``The covariant scattering and cohomology
of $W_3$ strings,'' preprint KCL-TH-93-2,
to be published in Int. J. Mod. Phys. {\bf A}
\item{[6]} H. Lu, C. N. Pope, S. Schrans and X-J. Wang, ``The
interacting $W_3$ string,'' CTP-TAMU-86/92.
\item{[7]} H. Lu, C. N. Pope, S. Schrans and X-J. Wang, ``On the spectrum
and scattering of $W_3$ strings,'' CTP-TAMU-4/93.
\item{[8]} J. Thierry-Mieg, Phys. Lett. {\bf B197} (1987) 368.
\item{[9]} S. Wolfram, ``Mathematica: a system for doing mathematics by
computer,'' Addison Wesley, 1991.
\item{[10]} K. Thielemans, Int. J. Mod. Phys. {\bf C2} (1991) 787.
\item{[11]} E. Bergshoeff, H. J. Boonstra, M. de Roo, S. Panda and
A. Sevrin, ``On the BRST operator of W-strings,'' preprint UG-2/93.
\item{[12]} A. B. Zamolodchikov and V. A. Fateev, Sov. Phys. JETP {\bf 62}
1985 215.
\item{[13]} I. Bakas and E. Kiritsis, Int. J. Mod. Phys. {\bf A7 [Suppl. 1A]}
(1992) 339.
F. Yu and Y-S Wu, Phys. Rev. Lett. {\bf 68} (1992) 2996.
\item{[14]} A Gerasimov, A. Marshakov and A. Morozov, Nucl. Phys. {\bf B328}
(1989) 664; J. Distler and Z. Qiu, Nucl. Phys. {\bf B336} (1990) 533;
K. Ito and Y. Kazama, Mod. Phys. Lett. {\bf A5} (1990) 215;
P. Griffin and O. Hern\'andez, Int. J. Mod. Phys. {\bf A7} (1992) 1233;
F. Narganes-Quijano, Int. J. Mod. Phys. {\bf A6} (1991) 2611.
\item{[15]} M. Freeman and P. West, in preparation
\item{[16]} C. Hull, ``New realisations of minimal models and the structure
of W-strings,'' NSF-ITP-93-65.

\end